\documentclass[aps,prb,twocolumn,showpacs]{revtex4}
\usepackage{amsmath}
\usepackage{amssymb}
\usepackage{graphicx}

\begin{document}
\title{Electric Field Induced Collapse of Charge-Ordered Phase in Manganites}
\author{S. Dong}
\email{saintdosnju@gmail.com}
\author{C. Zhu}
\author{Y. Wang}
\author{F. Yuan}
\author{K.F. Wang}
\author{J.-M. Liu}
\email{liujm@nju.edu.cn}
\affiliation{Nanjing National Laboratory of Microstructures, Nanjing University, Nanjing 210093, China\\
International Center for Materials Physics, Chinese Academy of Sciences,Shenyang, China}
\date{\today}

\begin{abstract}
The colossal electroresistance in manganites is a concomitant of the insulator-to-metal phase
transition induced by electric field. A phenomenological phase transition model is proposed to
study this electric field induced collapse of charge-ordered phase. The hysteresis of the phase
transition is well explained using the effective medium approximation. The volume fraction of
metallic region at the metal-to-insulator transition point is estimated as $30\%$. In addition, it
is found that the critical electric field to melt the charge-ordered phase decreases with the
applied magnetic field.
\end{abstract}

\maketitle

Manganites, a typical class of strongly correlated electron system, have been intensively studied
in the last decade, due to their unusual behaviors such as colossal magnetoresistance
(CMR).\cite{Dagotto:Bok} The existence of abundant phases in manganites, which are multiplicate in
macroscopic properties but close from one and another in free energy, not only is a challenge for basic physical research, but also allows an opportunity for potential applications. The
phase transition induced by a magnetic field, e.g. from a charge-ordered (CO) antiferromagnetic
(AFM) phase to a ferromagnetic (FM) phase, can cause an insulator-to-metal transition (IMT)
correspondingly.\cite{Yoshizawa:Prb, Dong:Jpcm} However, the utilization of the CMR effect is
unprosperous and an important embarrassment is that the required magnetic field is too large for
realization of magnetic storage. In addition, the IMT can be induced by many perturbations
other than magnetic field, e.g. hydrostatic pressure\cite{Moritomo:Prb} or substrate
strain,\cite{Xiong:Jap} electric field or current,\cite{Asamitsu:Nat, Ponnambalam:Apl, Rao:Prb,
Ma:Prb, Parashar:Jap, Garbarino:Prb} photon illumination of infrared laser,\cite{Fiebig:Sci} visual
light laser\cite{Smolyaninova:Apl} or X-ray.\cite{Kiryukhin:Nat} These phenomena open new
approaches for applications of manganites.

Compared to others, the transition switched by electric voltage or electric current may be more convenient for potential utilization. Many experimental studies on the electric effect argue that there are three main actions on the conduction: (1) The Joule self-heating effect can raise the local temperature ($T$) and change the resistivity ($\rho$) correspondingly since $\rho$ is $T$-dependent;\cite{Tokunaga:Prl, Yang:PRB, Sacanell:Jap, Mercone:Jap} (2) The interface between metal electrode and perovskite oxide may cause a polarity-dependent resistive switching under pulse electric field when two wires measurement is performed;\cite{Baikalov:Apl, Tsui:Apl} (3) The melting of CO state can give rise to a colossal negative electroresistance (CER), because the original CO state is insulated while the final FM phase is conductive.\cite{Rao:Prb, Ponnambalam:Apl, Garbarino:Prb} The magnitude of resistivity change in the CER effect is similar to that of CMR. In some cases, these actions may coexist and compete with each other.

The CER effect, which was first observed in Pr$_{0.7}$Ca$_{0.3}$MnO$_{3}$,\cite{Asamitsu:Nat} shows a first-order transition feature: an obvious hysteretic region charactered by two distinct critical electric voltages. The upper threshold for turning the high resistivity (HR) state to the low resistivity (LR) state is larger than the lower threshold which prevents the transition from the LR state to the HR state. The upper threshold voltage, which makes the CO phase to collapse, decreases with applied magnetic field and is minimized to zero when the magnetic field is strong enough to melt the CO state alone. These phenomena can also be found in other manganites with CO phase.\cite{Ma:Prb, Parashar:Jap, Garbarino:Prb} Some theoretical explanations have been proposed, e.g. the depinning of the randomly pinned charge solid.\cite{Rao:Prb} However, the current understanding of the CER effect remains insufficient. Especially, it is not clear in theory whether the CO collapse is induced by electric field or current. And when will the CO state collapse and rebuild remains unpredictable. In this Brief Report, the CER effect will be studied using a phenomenological phase transition model based on the idea of phase separation (PS) and dielectric breakdown.\cite{Asamitsu:Nat, Ponnambalam:Apl}

In our model, a bulk CO phase manganite is submitted to a homogenous electric field $E$, as shown in Fig. 1, step 1. The relative dc dielectric constant of this CO phase is $\varepsilon_{r}$. The energy of the CO phase is lower than the FM phase in zero field. The energy gap between them can be simply estimated as $\mu_{0}H_{c}M_{s}$, because the material can gain Zeeman energy from the magnetic field  for the FM spin arrangement, but none for the AFM spin arrangement. Here $\mu_{0}$ is the magnetic permeability of vacuum, $\mu_{0}H_{c}$ is the critical value of magnetic field (in unit of Tesla) to break the CO state and $M_{s}$ is the saturated magnetization of the FM state. Considering the FM state of manganites is half-metallic with almost $100\%$ polarization of $3d$ electrons, $M_{s}$ (per mol) equals $N_{A}g_{L}\mu_{B}S$, where $N_{A}$ is the Avogadro's number,
$g_{L}$ is the Lande factor of spin, $\mu_{B}$ is the Bohr magneton and $S$ is the spin momentum of Mn $3d$ electrons in unit of the Plank constant $\hbar$. Besides magnetic field, electric field can also modulate the system energy. For the AFM CO insulator, charges are mainly localized and dielectric polarization (\textbf{P}) can be estimated as $\varepsilon_{0}\varepsilon_{r}\textbf{E}$, where $\varepsilon_{0}$ is the permittivity of free space. This polarization lowers the energy per volume by $\textbf{P}\cdot\textbf{E}$. In contrast, in the FM metal region, the carriers are itinerant and can assemble on the surface driven by electric field. These surface charges build an inner field opposing to the external field $E$, as shown in Fig. 1, step 2. The existence of the FM region reduces the thickness of the CO region and raises the electric field in the CO phase. This effect leads to the decline of the polarization energy by $\varepsilon_{0}\varepsilon_{r}E^{2}V$, as further proved in the Appendix, where $V$ is the FM volume. Thus, if  a FM region embedded in the CO matrix can be generated, the critical electric field $E_{c}$ should be enough to fill the energy gap between the CO/FM phase:
\begin{equation}\label{eq:Eq1}
\varepsilon_{0}\varepsilon_{r}E_{c}^{2}V=N_{A}\mu_{0}H_{c}g_{L}\mu_{B}S,
\end{equation}
where the quantity of FM phase has been normalized.

\begin{figure}
\includegraphics[width=200pt]{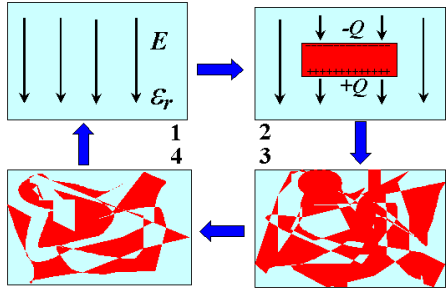}
\caption{(Color online) A full cycle of the collapse and rebuilding of CO state in electric field. The red part indicates the conductive FM region while the cyan background stands for the insulated CO AFM matrix. The $1$st/$3$st step is the high/low resistivity state, while the $2$st/$4$st step is the upper/lower critical point, respectively.}
\end{figure}

It can be obtained from the above equation: $E_{c}\sim\sqrt{H_{c}}$, a relationship between the critical electric field and critical magnetic fields to melt the CO phase, which is different from the $E_{c}\sim H_{c}$ estimated by Sacanell \textit{et al}.\cite{Sacanell:Jap} All parameters in Eq. (\ref{eq:Eq1}) can be measured on the same
conditions. When the electric field $E$ is below $E_{c}$, the CO phase is stable against the FM transition. The material keeps poor conductive and is less susceptible to the increasing electric field. However, once $E$ is beyond $E_{c}$, some regions in the material will turn to be FM metal. The electric field on the rest insulated region will be enhanced because the effective thickness of the CO dielectric is decreased. It is a positive feedback process that induces a collapse of the CO phase. Consequently, the percolative conductive paths run through the bulk material, as shown in Fig. 1, step $3$. The system turns to be the LR state and the remnant CO regions are in short circuit state. The whole process is sketched in Fig. 1, step 1-2-3, and the corresponding relationship between $\rho$ and $E$ is sketched as the curve 1-2-3 in Fig. 2. These consequences of our model are consistent with the experimental observations. In experiments, the resistivity has a abrupt colossal drop when the applied voltage is over a threshold, while below the threshold the conductive is almost independent of the voltage.\cite{Asamitsu:Nat}

\begin{figure}
\includegraphics[width=200pt]{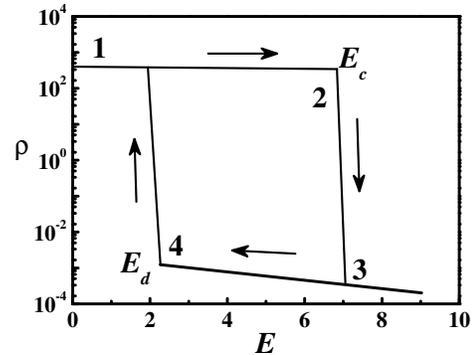}
\caption{Sketch of resistivity change in a full cycle as function of electric field, corresponding to the $4$ steps in Fig. 1. The scale of axis is just a guide to eyes.}
\end{figure}

Equation (\ref{eq:Eq1}) can be extended to explain more phenomena in CER. On one hand, the hysteretic feature is obvious in the electric field induced first-order phase transition. There are two electric field thresholds in a full cycle, which construct an approximate rectangular loop in the $E$-$\rho$ diagram, as shown in Fig. 2. The
two critical values correspond to the first-order CO-to-FM/FM-to-CO transitions, respectively. The upper threshold $E_{c}$ for the IMT transition has already been derived. In the following, the other metal-to-insulator transition (MIT), which occurs when $E$ is turned down to the lower threshold $E_{d}$, will be investigated. Since the LR state is an inhomogeneous state with percolating character, the MIT point should be a terminal of the percolation, as shown in Fig. 1, step 4-1. Here, an effective medium approximation is used to approach the phase coexistence system. Eq. (\ref{eq:Eq1}) is used once more, with $\varepsilon_{r}$ replaced by an effective dielectric constant $\varepsilon_{e}$. It is reasonable that the existence of metallic regions increases the electric field in the insulated regions and raises the macroscopic dielectric constant, as mentioned before. With this substitution, a relationship is obtained:
\begin{equation}\label{eq:Eq2}
\frac{E_{c}}{E_{d}}=\sqrt{\frac{\varepsilon_{e}}{\varepsilon_{r}}}.
\end{equation}
In Asamitsu \textit{et al}'s experiment on Pr$_{0.7}$Ca$_{0.3}$MnO$_{3}$ single crystal, the upper threshold of
voltage is about $750$ V and the lower one is about $250$ V (measured at $20$ K with $0.99$ mm span between electrodes).\cite{Asamitsu:Nat} The ratio $E_{c}/E_{d}$ is $3$. In other word, $\varepsilon_{e}$ is about $9\varepsilon_{r}$. According to the effective medium theory, the effective dielectric constant of a two-component mixture system is:\cite{Nan:Pms}
\begin{equation}\label{eq:Eq3}
\varepsilon_{e}=\frac{1}{4}(P+\sqrt{8\varepsilon_{A}\varepsilon_{B}+P^2)},
\end{equation}
where $P$ is defined as $(3c-1)\varepsilon_{A}-(3c-2)\varepsilon_{B}$, $\varepsilon_{A}$ is the dielectric constant of one component with volume fraction $c$ and $\varepsilon_{B}$ is for the other component with volume fraction ($1-c$). Here, $\varepsilon_{A}$ of the metal component is far larger than $\varepsilon_{B}$ ($=\varepsilon_{r}$) of the CO component. By substituting $\varepsilon_{B}/\varepsilon_{A}\sim0$ and
$\varepsilon_{e}=9\varepsilon_{B}$ into Eq. (\ref{eq:Eq3}), the metal volume fraction $c$ is found to be about $30\%$ on the edge of the MIT occurs  (point $4$ in Fig. 2). This fraction is only slightly smaller than the percolative threshold (point $3$ in Fig. 2) of the effective medium theory in three-dimensional systems, $1/3$. In experiments, there is a raise of resistivity before the MIT, responding to the slight decrease of metal volume fraction (from $1/3$ to $30\%$), as sketched in Fig. 2, curve $3-4$. When $E$ is below $E_{d}$, the metal state becomes unstable and turns back to the CO state. This first-order phase transition is also a positive feedback process. In addition, if the relation between $c$ and $E$ can be found otherwise, the lower threshold $E_{d}$ can be obtained from Eq. (\ref{eq:Eq2}-\ref{eq:Eq3}) by reversal derivation. On the other hand, experimentally, the critical value $E_{c}$ decreases with applied magnetic field $H$, till $E_{c}=0$ when $H$ is beyond the threshold value $H_{c}$. This effect can be attributed to the magnetic energy contribution $\mu_{0}HM_{s}$, which supplies part of the penalty energy in the CO-FM phase transition. Therefore, with external magnetic field, $H_{c}$ in Eq. (\ref{eq:Eq1}) should be substituted by ($H_{c}-H$). The critical electric field becomes:
\begin{equation}\label{eq:Eq4}
E_{c}(H)=E_{c}(0)\sqrt{1-H/H_{c}},
\end{equation}
as shown in Fig. 3. Here $E_{c}(0)$ derived from Eq. (\ref{eq:Eq1}) is the critical electric field in zero magnetic field. Eq. (\ref{eq:Eq4}) is in good agreement with the experimental data of Pr$_{0.7}$Ca$_{0.3}$MnO$_{3}$, in which the voltage causing the CO collapse decreases from $750$ V (zero magnetic field) to zero ($\mu_{0}H_{c}\simeq4$ Tesla) in the relationship analogous to the curve in Fig. 3.\cite{Asamitsu:Nat}

\begin{figure}
\includegraphics[width=200pt]{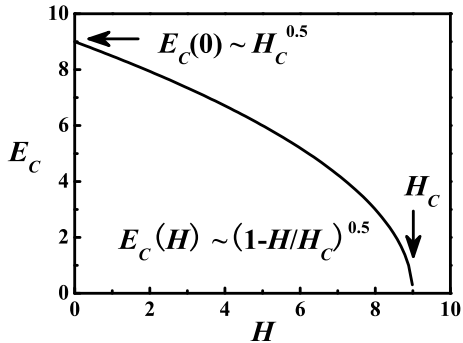}
\caption{The critical electric field as function of applied magnetic field. The scale of axis is just a guide to eyes.}
\end{figure}

There is an argument that the CER might be induced by electric current too. It is possible when the polarized carriers in the FM region are pumped into neighboring CO region. Then the AFM CO insulator is polarized to the FM metal. However, this idea is incomplete, because the polarized carrier may rebound instead of immission at the interface of AFM region when the driving field ($E$) is not strong enough. Our work just reveals the threshold $E_{c}$, only beyond which the current can flow though the CO region.

At last, the $E_{c}$ estimated from our model is compared with the experimental result on Pr$_{0.7}$Ca$_{0.3}$MnO$_{3}$. Using the data $H_{c}\sim4$ Tesla\cite{Asamitsu:Nat} and $\varepsilon_{r}\sim80$\cite{Freitas:Prb}at $20$ K, the calculated $E_{c}$ is $\sim10^7$ V/m, while the experimental value is $\sim10^6$ V/m.\cite{Asamitsu:Nat} It is seemed that the calculated $E_{c}$ is compatible to the measured one on the same order of magnitude. In fact, considering the simplicity of the model, the quantitative difference between them is not remarkable, which may be ascribed to the following effects: (1) In experiments, the leakage current can heat the material and raise the local $T$. In the work of Asamitsu \emph{et al}, $T$ jumped from $20$ K to $25$ K when the CO collapse occurred. Because $\varepsilon_{r}$ is strongly dependent on $T$,\cite{Freitas:Prb} the increase of $\varepsilon_{r}$ due to $T$ rising will reduce the $E_{c}$. In addition, $H_{c}$ is also dependent of $T$;\cite{Yoshizawa:Prb} (2) Pr$_{0.7}$Ca$_{0.3}$MnO$_{3}$ is in a spin canted AFM phase rather than a ideal AFM state at low $T$.\cite{Yoshizawa:Prb} In this case, the energy difference between the CO/FM phases would be less than $\mu_{0}H_{c}M_{s}$ because of the nonzero magnetic moment in the CO phase; (3) Due to the anomalous shapes of FM clusters, the local electric field may be somehow larger than the average value at the edge of phase interface, where point discharge becomes inevitable. Due to its positive feedback feature, the CO will collapse in advance with smaller $E_{c}$. In consideration of all of those
effects if possible, a much better consistence between the calculated and measured values of $E_{c}$ may yield.  For comparison, the $E_{c}\sim H_{c}$ relation given in Ref. 17 can be also used to estimate the critical electric field, which is about $10^{5}$ V/m for Pr$_{0.7}$Ca$_{0.3}$MnO$_{3}$. This field is smaller than the real value in this case. This relation may be more suitable for the LaPrCaMnO series in which PS are natural even without field.  In the case of PrCaMnO, dielectric breakdown may be the main factor in the CER.\cite{Asamitsu:Nat, Ponnambalam:Apl} Certainly, the effect of the electric field on the CO state in manganites is more complicated and various behaviors other than the collapse are also displayed.\cite{Mercone:Jap} The present model grasps the main physical ingredients of the high electric field induced CO phase collapse which is one of the most prominent effects in manganites.

In conclusion, we have proposed a phenomenological phase transition model to study the electric field induced collapse of charge-ordered phase in manganites. The upper and lower critical values for the hysteretic region of electric field induced phase transitions are well explained using the effective medium approximation. In addition, the relationship between threshold of electric field and applied magnetic field is also derived.

This work was supported by the Natural Science Foundation of China (50332020, 10021001) and National Key Projects for Basic Research of China (2002CB613303, 2006CB0L1002).

\section{APPENDIX}
For simplification, the AFM CO dielectric is assumed to be an unit area, $d$ thick cuboid, as shown in Fig. 1, step 1. The applied voltage between the upper and lower surfaces is $U$. Therefore, the uniform electric field $E$ is $U/d$ and the polarization energy is:
\begin{equation}\label{eq:Eq5}
E_{P}=-\varepsilon_{0}\varepsilon_{r}E^{2}d.
\end{equation}
When an unit-area, $d$ thick FM slab is inserted into the CO medium, the thickness of the CO dielectric becomes $d-a$, as shown in Fig. 1, step 2. Here the dielectric constant of the FM phase ($\varepsilon_{m}$) is assumed to be far larger than $\varepsilon_{r}$. Thus the electric field in the FM slab is near zero. Consequently, the electric field in the CO phase above and below the FM slab is raised to $U/(d-a)$. The polarization energy is: \begin{equation}\label{eq:Eq6}
E_{P}^{'}=-\varepsilon_{0}\varepsilon_{r}(\frac{U}{d-a})^{2}(d-a)=E_{P}+E_{P}\frac{a}{d-a}.
\end{equation}
Since the FM slab is very small ($a\ll d$, so $d\sim d-a$), the energy reduction is $\varepsilon_{0}\varepsilon_{r}E^{2}a$. Here the polarization energy in the FM slab is negligible because the electric field in the FM slab is in proportion to $\varepsilon_{r}/\varepsilon_{m}$, which is near zero.

\bibliographystyle{apsrev}
\bibliography{ref}
\end{document}